\documentclass[%
 aip,
 amsmath,amssymb,
preprint,%
 fleqn,
]{revtex4-1}

\usepackage{graphicx}
\usepackage{dcolumn}
\usepackage{bm}

\begin{document}


\title{Dust acoustic solitary structures in presence of nonthermal ions, isothermally distributed electrons and positrons}

\author{Ashesh Paul}
\affiliation{ Department of Mathematics, Jadavpur University, Kolkata
- 700032, India.}%
\author{Anup Bandyopadhyay}
\email{abandyopadhyay1965@gmail.com}%
\affiliation{ Department of Mathematics, Jadavpur University, Kolkata
- 700032, India.}%
\date{\today}

\begin{abstract}
\noindent Arbitrary amplitude dust acoustic solitary structures have been investigated in a four component multi-species plasma consisting of negatively charged dust grains, nonthermal ions, isothermally distributed electrons and positrons including the effect of dust temperature. We have used the Sagdeev pseudo-potential method to discuss the arbitrary amplitude steady state dust acoustic solitary structures in the present plasma system. We have designed a computational scheme to draw the existence domains of different dust acoustic solitary structures. We have observed only negative potential solitary waves for isothermal ions. But for strong nonthermality of ions the system supports positive potential solitary waves, positive potential double layers and coexistence of solitary waves of both polarities. The positive potential solitary waves are restricted by the positive potential double layers but negative potential double layer has not been found for any parameter regime. The system does not support dust acoustic supersoliton of any polarity. The concentration of positrons plays an important role in the formation of positive potential double layers. Finally, the phase portraits of the dynamical system have been presented to confirm the existence of different dust acoustic solitary structures.   
\end{abstract}

\maketitle

\section{\label{sec:intro5}Introduction}

It is well established in the literature that the presence of highly charged dust grains having large masses introduces several new features on the behaviour of different nonlinear waves.
Considering the dynamics of negatively charged dust particles in a background of hot electrons and ions, Rao \textit{et al.} \cite{rao90} first reported the existence of low frequency dust acoustic (DA) solitons of both polarities. Later, DA solitary structures have been observed in different laboratory experiments \cite{barkan95,praburam96,pieper96,thompson97,bandyopadhyay08} which support the theoretical prediction of Rao \textit{et al.} \cite{rao90} on the existence of the DA waves. It is also predicted that DA waves may be present in various dusty astrophysical sites such as in the planetary rings \cite{melandso93,havnes95}, especially in Saturn's A and B rings \cite{yaroshenko07}, nebulon structures or the cloudlike structure observed in the Milky Way \cite{tian05}, in the vicinity of the lunar terminator \cite{popel13}, cometary tails \cite{mendis13,arshad14} etc. 

In cosmic plasma, the velocity distribution of particles may not always follow the Maxwell-Boltzmann velocity distribution. Rather, the charged particles are often nonthermal and follow non-Maxwellian velocity distribution functions due to the presence of an amount of high energetic particles which are not in thermodynamic equilibrium. The presence of such nonthermal energetic particles have been detected in various dusty astrophysical environments \cite{asbridge68,feldman83,lundin89,verheest00,shukla02,futaana03}. Although the actual mechanism for the formation of energetic particles in space plasma is yet to be an open problem, different non-Maxwellian distributions have been modelled in phase space to describe the behaviour of the energetic particles. The Cairns \cite{cairns95} nonthermal velocity distribution function is one of the widely used non-Maxwellian velocity distribution function. A number of authors \cite{mamun96DA,mendoza00,maharaj04,kourakis05,maharaj06,verheest08,verheest09} investigated DA solitary structures in dusty plasmas where they considered Cairns \cite{cairns95} model for ions and/or electrons. Das \textit{et al.} \cite{das09} investigated DA solitary waves and double layers in a plasma consisting of negatively charged static dust grains, Cairns \cite{cairns95} distributed nonthermal ions and isothermal electrons including the effect of dust temperature. They reported the existence of positive potential solitary waves (PPSWs), negative potential solitary waves (NPSWs), coexistence of solitary waves of both polarities and most importantly, the existence of positive potential double layers (PPDLs).

Positron, being an antimatter of electron can be considered as lighter species of a plasma and may acquire energy as high as electron. Positrons are the most common antimatter found in astrophysical plasma and easy to produce in the laboratory. In our previous papers \cite{paul2016,paul17ppr,paul17pop}, we have clearly stated that the most of the astrophysical plasmas contain highly charged dust grains, and consequently the four component electron-positron-ion-dust (e-p-i-d) plasma may be found in numerous cosmic environments. For example, in and around the pulsars \cite{shukla04}, near the surface of the neutron stars \cite{zeldovich1971,shukla04}, in the hot spots on dust ring in the galactic centre \cite{zurek1985}, interstellar medium \cite{zurek1985,higdon09,shukla2008}, interior regions of accretion disks near neutron stars and magnetars \cite{dubinov12}, in Milky way \cite{shukla2008}, in the magnetosphere and in the ionosphere of the Earth \cite{alfven1981,gusev2000,gusev2001}, in the magnetosphere of the Jupiter \cite{merlino2006} and the Saturn \cite{horanyi2004} etc. Apart from these astrophysical plasmas, e-p-i-d plasma can be produced in laboratory environments \cite{shukla04,dubinov12} also.


Using the reductive perturbation method Jehan \textit{et al.} \cite{jehan09} investigated the planar and non-planar DA solitary waves in an unmagnetized e-p-i-d plasma whose constituents are negatively charged dust and Boltzmann distributed electrons, positrons and ions. Wang and Zhang \cite{wang15PRAMANA,wang14} investigated small but finite amplitude DA solitary waves in a planar unmagnetized dusty plasma consisting of electrons, positrons, ions and negatively charged dust particles with different sizes and masses. Using the Sagdeev \cite{sagdeev66} pseudo-potential approach Esfandyari-Kalejahi \textit{et al.} \cite{esfandyari12} studied DA waves in a four component dusty plasma with negatively charged inertial dust grains and Boltzmann distributed electrons, positrons, and ions. They reported that the system only supports negative potential solitary waves and it does not support double layer of any polarity.  Again, employing the Sagdeev \cite{sagdeev66} pseudo-potential technique DA solitary structures have been investigated by Saberian \textit{et al.} \cite{saberian17} in a plasma consisting of superthermal electrons, positrons, and positive ions in presence of negatively charged dust grains with finite-temperature. They reported the existence of negative potential solitary waves only.

Therefore, from the paper of Das \textit{et al.} \cite{das09} we see that a three component electron-ion-dust (e-i-d) plasma supports DA double layer of positive potential when the velocity distribution functions of ion species are considered nonthermal due to Cairns \textit{et al.} \cite{cairns95}. In fact, the system supports DA double layer of positive polarity when the nonthermality of ions exceeds a certain critical value. In the present investigation, our main aim is to investigate the nature of existence of different DA solitary structures when an amount of positron is injected in the same plasma system of Das \textit{et al.} \cite{das09}. For that purpose we extend the plasma model of Das \textit{et al.} \cite{das09} into a four component e-p-i-d plasma consisting of negatively charged dust grains, Cairns \cite{cairns95} distributed nonthermal ions, isothermally distributed electrons and positrons. The Cairns \cite{cairns95} model for nonthermal velocity distribution of ions has not been considered earlier in the investigations of DA waves in four component e-p-i-d plasma.

In the present paper, the investigation on DA solitary structures in a four component e-p-i-d plasma has been carried out on the following directions.
\begin{itemize}
	\item The equation of pressure for dust fluid is taken into account to include the effect of dust temperature. 
	\item We have considered Cairns \cite{cairns95} distributed nonthermal ions. It has been observed that for isothermal ions the system only supports NPSWs. This result completely agrees with the findings of Esfandyari-Kalejahi \textit{et al.} \cite{esfandyari12}. But in the present paper, we have found that for strong nonthermality of ions the system supports PPDLs as well as PPSWs and coexistence of solitary waves of both polarities.
	\item To investigate the existence of different DA solitary structures in the present plasma system we have considered the qualitatively different existence domains or the compositional parameter spaces which describe a complete scenario of the nature of existence of the solitary structures with respect to any parameter of the plasma system.
	\item For the first time, we have introduced the phase portrait analysis of the dynamical system corresponding to the DA solitary structures supported by the system.
\end{itemize}

\section{Basic Equations \& Energy Integral}\label{sec:basic_eq_c5}

The following are the governing equations describing the non-linear behaviour of dust acoustic waves propagating along x-axis in a collisionless unmagnetized  dusty plasma consisting of negatively charged warm dust grains, Cairns \cite{cairns95} distributed nonthermal ions and isothermally distributed electrons and positrons:
\begin{eqnarray}\label{continuity}
\frac{\partial n_{d}}{\partial t}+\frac{\partial}{\partial x}(n_{d}u_{d})=0,
\end{eqnarray}
\begin{eqnarray}\label{momentum}
M_{sd}^{2}\bigg(\frac{\partial u_{d}}{\partial t}+u_{d}\frac{\partial u_{d}}{\partial x}\bigg)+\frac{(1-\mu)\sigma_{di}}{Z_{d}n_{d}}\frac{\partial p_{d}}{\partial x}-\frac{\partial \phi}{\partial x}=0,
\end{eqnarray}
\begin{eqnarray}\label{pressure}
\frac{\partial p_{d}}{\partial t}+u_{d}\frac{\partial p_{d}}{\partial x}+\gamma p_{d} \frac{\partial u_{d}}{\partial x}=0,
\end{eqnarray}
\begin{eqnarray}\label{poisson}
\frac{\partial^{2} \phi}{\partial x^{2}}=-\frac{M_{sd}^{2}-\gamma \sigma_{di}}{\sigma_{ie}^{2}(1-\mu)}\bigg(n_{i}-n_{e}+n_{p}-Z_{d}n_{d}\bigg).
\end{eqnarray}
Here $n_{d}$, $n_{i}$, $n_{e}$, $n_{p}$, $u_{i}$, $p_{i}$, $\phi$, $x$ and $t$ are, respectively, the dust particle number density, the ion number density, the electron number density, the positron number density, velocity of dust fluid, dust fluid pressure, electrostatic potential, spatial variable and time, and these have been normalized by $n_{0}$ ($=n_{i0}+n_{p0}=n_{e0}+Z_{d}n_{d0}$), $n_{0}$, $n_{0}$, $C_{Dd}$ (linearized velocity of the DA wave in the present plasma system for long-wavelength plane wave perturbation), $n_{d0}K_{B}T_{d}$, $\Phi=\frac{K_{B}T_{i}}{e}$, $ \lambda_{Dd} $ (Debye length of the present plasma system) and $\lambda_{Dd}/C_{Dd}$ with $n_{e0}$, $n_{i0}$, $n_{p0}$ and $n_{d0}$ are, respectively, the equilibrium number densities of electrons, ions, positrons and dust particulates, $ \gamma(=3) $ is the adiabatic index, $ Z_{d} $ is the number of electrons residing on a dust grain surface, $-e$ is the charge of an electron, $T_{d}$ ($T_{i}$) is the average temperature of dust grains (ions), and $K_{B}$ is the Boltzmann constant.

The normalized number densities of nonthermal ions, isothermal electrons and isothermal positrons are given by
\begin{eqnarray}\label{ni}
n_{i} = (1-p)(1+\beta_{i}\phi+\beta_{i}\phi^{2})e^{-\phi},
\end{eqnarray}
\begin{eqnarray}\label{ne}
n_{e} = \mu e^{\sigma_{ie} \phi}.
\end{eqnarray}
\begin{eqnarray}\label{np}
n_{p} = p e^{-\sigma_{ip} \phi}.
\end{eqnarray}
The above equations are supplemented by the following unperturbed charge neutrality condition
\begin{eqnarray}
n_{i0}+n_{p0}=n_{e0}+Z_{d}n_{d0}.
\end{eqnarray} 

The expressions of $M_{sd}$ and the basic parameters $p$, $\mu$, $\sigma_{di}$, $\sigma_{ie}$, $\sigma_{pe}$, $\sigma_{ip}$ are given by the following equations: 
\begin{eqnarray}\label{Ms}
M_{sd}=\sqrt{\gamma\sigma_{di}+\frac{(1-\mu)}{(1-p)(1-\beta_{i})+\mu \sigma_{ie}+p \sigma_{ip}}},	
\end{eqnarray}
\begin{eqnarray}\label{p}
p=\frac{n_{p0}}{n_{0}},~\mu=\frac{n_{e0}}{n_{0}},~\sigma_{di}=\frac{T_{d}}{Z_{d}T_{i}},~\sigma_{ie}=\frac{T_{i}}{T_{e}},~\sigma_{pe}=\frac{T_{p}}{T_{e}},~\sigma_{ip}=\frac{T_{i}}{T_{p}},	
\end{eqnarray}
where $T_{e}$ ($T_{p}$) is the average temperature of electrons (positrons) and $\beta_{i}$ is the nonthermal parameter associated with the Cairns model \cite{cairns95} for ion species, and according to  Verheest \& Pillay \cite{verheest08}, the physically admissible bounds of $\beta_{i}$ is given by $0 \leq \beta_{i} \leq \frac{4}{7} \approx 0.6$.

In fact, the linear dispersion relation of the DA wave for the present dusty plasma system can be written as
\begin{eqnarray}\label{dispersion_relation}
\frac{\omega}{k} &=& C_{Dd}\sqrt{\frac{1+\frac{\gamma\sigma_{di}}{M_{sd}^{2}}k^{2}\lambda_{Dd}^{2}}{1+k^{2}\lambda_{Dd}^{2}}},\\ C_{Dd} &=& C_{d}M_{sd},C_{d}=\sqrt{\frac{Z_{d}K_{B}T_{i}}{m_{d}}},
\end{eqnarray}
where 
\begin{eqnarray}\label{lambda_D_square}
\frac{1}{\lambda_{Dd}^{2}}=\frac{1-\beta_{i}}{\lambda_{Di}^{2}}+\frac{1}{\lambda_{Dp}^{2}}+\frac{1}{\lambda_{De}^{2}}~,
\end{eqnarray}
\begin{eqnarray}\label{lambda_Di}
\lambda_{Di}^{2}=\frac{K_{B}T_{i}}{4\pi e^{2}n_{i0}}~,~
\lambda_{Dp}^{2}=\frac{K_{B}T_{p}}{4\pi e^{2}n_{p0}}~,~
\lambda_{De}^{2}=\frac{K_{B}T_{e}}{4\pi e^{2}n_{e0}}.
\end{eqnarray}
Here, $\omega$ and $k$ are, respectively, the wave frequency and wave number of the plane wave perturbation and $m_{d}$ is the mass of a dust grain. For long-wave length plane wave perturbation, i.e., for $ k \rightarrow 0 $, from linear dispersion relation (\ref{dispersion_relation}), we have,
\begin{eqnarray}
\lim_{k \to 0}\frac{\omega}{k} =C_{Dd} \mbox{   and   } \lim_{k \to 0}\frac{d\omega}{dk} = C_{Dd}
\end{eqnarray}
and consequently the dispersion relation (\ref{dispersion_relation}) shows that the linearized velocity of the DA wave for long-wave length plane wave perturbation is $ C_{Dd} $ with $ \lambda_{Dd} $ as the Debye length. So, according to the prescription of Dubinov \cite{dubinov09a}, here each spatial coordinate is normalized by $\lambda_{Dd}$ and the time is normalized by $\frac{\lambda_{Dd}}{C_{Dd}}$.

To investigate the steady state arbitrary amplitude DA solitary structures, we make all the dependent variables depend only on a single variable $ \xi=x-Mt $ where $M$ is the dimensionless velocity of the wave frame normalized by the linearized DA speed ($C_{Dd}$) for long-wavelength plane wave perturbation. Using this transformation and applying the boundary conditions:\\ $ \big(n_{d},p_{d},u_{d},\phi,\frac{d\phi}{d\xi}\big)\rightarrow \big(\frac{1-\mu}{Z_{d}},1,0,0,0\big)\mbox{    as    }  |\xi|\rightarrow \infty,
$\\ we get the following energy integral:
\begin{eqnarray}\label{energy_integral}
\frac{1}{2}\bigg(\frac{d\phi}{d\xi}\bigg)^{2}+V(\phi)=0,
\end{eqnarray}
where
\begin{eqnarray}\label{V_phi_1}
V(\phi) = (M_{sd}^{2}-3\sigma_{di}) \Big[~V_{d} -\frac{\mu }{1-\mu} \frac{1 }{\sigma_{ie}} V_{e}+\frac{1-p}{1-\mu}V_{i}+\frac{p}{1-\mu}\frac{1}{\sigma_{ip}}V_{p}\Big],
\end{eqnarray}
\begin{eqnarray}\label{V_d_1}
V_{d} = M^{2}M_{sd}^{2}+\sigma_{di} -N_{d}\Big[M^{2}M_{sd}^{2}+3\sigma_{di}+2\phi -2\sigma_{di}N_{d}^{2}\Big],
\end{eqnarray}
\begin{eqnarray}\label{N_d_1}
N_{d}=\frac{Z_{d}n_{d}}{1-\mu}=\frac{MM_{s}\sqrt{2}}{(\sqrt{\phi-\Phi_{M}}+\sqrt{\phi-\Psi_{M}})},
\end{eqnarray}
\begin{eqnarray}\label{Phi_M_1}
\Phi_{M} &=& -\frac{1}{2}\Big(MM_{s}+\sqrt{3\sigma_{di}}\Big)^{2},
\end{eqnarray}
\begin{eqnarray}\label{Psi_M_1}
\Psi_{M} &=& -\frac{1}{2}\Big(MM_{s}-\sqrt{3\sigma_{di}}\Big)^{2},
\end{eqnarray}
\begin{eqnarray}\label{V_i_1}
V_{i} &=& (1+3\beta_{i})-\big(1+3\beta_{i}+3\beta_{i}\phi+\beta_{i}\phi^{2}\big)e^{-\phi},
\end{eqnarray}
\begin{eqnarray}
V_{e} &=& e^{\sigma_{ie} \phi}-1,
\end{eqnarray}
\begin{eqnarray}
 V_{p} &=& 1-e^{-\sigma_{ip} \phi}.
\end{eqnarray}
For the existence of PPSW (NPSW), we have\\
	(i) $V(0)=V'(0)=0$ and $V''(0)<0$,\\
	(ii) $V(\phi_{m}) = 0$, $V'(\phi_{m}) > 0$ ($V'(\phi_{m}) < 0$) for some $\phi_{m} > 0$ ($\phi_{m} < 0$),\\
	(iii) $V(\phi) < 0$ for all $0 <\phi < \phi_{m}$ ($\phi_{m} < \phi < 0$).
	
For the existence of PPDL (NPDL), the condition (ii) is replaced by $V(\phi_{m}) = 0$, $V'(\phi_{m}) = 0$, $V''(\phi_{m}) < 0$ for some $\phi_{m} > 0$ ($\phi_{m} < 0)$. 

The condition (i) gives $M>M_{c}=1$, i.e., the solitary structures start to exist just above the curve $M = M_{c}=1$.

To define $ N_{d} $ as a real quantity, we must have $ \phi \geq \Psi_{M} $. This condition gives the upper bound of the Mach number $M$ for the existence of NPSWs. If we denote this upper bound as $M_{max}$, then $ M_{max} $ is the largest positive root of the equation $ V(\Psi_{M}) = 0 $ for the unknown $M$ (whenever the values of other parameters are specified) along with the condition $ V(\Psi_{M}) \geq 0 $ for all $ M \leq M_{max} $. So, $M$ assumes its upper limit $ M_{max} $ for the existence of all NPSWs when $\phi$ tends to $\Psi_{M}$, i.e., when dust density reaches to highest compression.

Following the construction of dust ion acoustic (DIA) double layers of any polarity as given in section 5.3 of Das \textit{et al.} \cite{das12}, we find that if the PPDL solution of the energy integral (\ref{energy_integral}) having amplitude $|\phi_{PPDL}|(=\phi_{PPDL}>0)$ exists at $M=M_{PPDL}$, then $M_{PPDL}$ is given by the following sequence of equations:
\begin{eqnarray}\label{M_ppdl}
M_{PPDL} = \frac{\sqrt{h(\phi_{PPDL})}}{M_{sd}},
\end{eqnarray}
where
\begin{eqnarray}\label{h_phi}
h(\phi) = \frac{S_{1}-\sigma_{di} -\frac{dS_{1}}{d\phi}\big[3\sigma_{di}+2\phi-2\sigma_{di}\big(\frac{dS_{1}}{d\phi}\big)^{2}\big]}{1+\frac{dS_{1}}{d\phi}},
\end{eqnarray}
\begin{eqnarray}\label{S}
S_{1} = \frac{\mu }{1-\mu} \frac{1}{\sigma_{ie}} V_{e}-\frac{1-p}{1-\mu}V_{i}-\frac{p}{1-\mu}\frac{1}{\sigma_{ip}}V_{p}.
\end{eqnarray}
Here $\phi_{PPDL}$ is the strictly positive root of the equation
\begin{eqnarray}\label{eta_phi_ppdl_npdl}
\eta(\phi_{PPDL}) = 0,
\end{eqnarray}
satisfying the following three conditions
\begin{eqnarray}\label{inequality_7}
h(\phi_{PPDL})-M_{c}^{2} > 0,
\end{eqnarray}
\begin{eqnarray}\label{inequality_8}
\phi_{PPDL}-g_{-}(\phi_{PPDL}) \geq 0,
\end{eqnarray}
\begin{eqnarray}\label{inequality_9}
\frac{d\eta}{d\phi}\bigg|_{\phi=\phi_{PPDL}} < 0,
\end{eqnarray}
where $\eta(\phi)$ and $g_{\pm}(\phi)$ are given by
\begin{eqnarray}\label{eta_phi}
\eta(\phi) = \frac{\sqrt{2h(\phi)}}{\sqrt{\phi-g_{+}(\phi)}+\sqrt{\phi}-g_{-}(\phi)}+\frac{dS_{1}}{d\phi},
\end{eqnarray}
\begin{eqnarray}\label{g_plus_minus_phi}
g_{\pm}(\phi) = -\frac{1}{2}\big(\sqrt{h(\phi)}\pm\sqrt{3\sigma_{di}}\big)^{2}.
\end{eqnarray}
The condition (\ref{inequality_8}), have been derived from the following restriction of $ \phi $ : $ \phi \geq \Psi_{M} = -\frac{1}{2}(MM_{s}-\sqrt{3\sigma_{di}})^{2} $. The similar analysis as discussed here is valid to find the Mach number ($M_{NPDL}$) corresponding to a NPDL solution of the energy integral (\ref{energy_integral}).

Now we are in a position to draw the existence domains with respect to any parameter of the system to discuss the nature of existence of different DA solitary structures supported by the present plasma model. In the next section, we have investigated the nature of existence of different DA solitary structures associated with the different solutions of the energy integral (\ref{energy_integral}) with the help of the qualitatively different existence domains.

\section{Existence Domains}\label{sec:existence_domain_c5}

The basic parameters of the present e-p-i-d plasma system are $p$, $\mu$, $\sigma_{di}$, $\sigma_{ie}$, $\sigma_{pe}$ and $\beta_{i}$. For the present dusty plasma system, we take $\sigma_{ie}=\sigma_{pe}=0.9$ and $\sigma_{di}=0.0001$. In the present paper, we will discuss the qualitatively different existence domains of the different solitary structures with respect to $\beta_{i}$ for different values of $p$ and $\mu$. To interpret the existence domains we have made a general description as follows:
\begin{itemize}
  \item Solitary structures start to exist just above the lower curve $ M = M_{c} =1 $.
  \item At each point on the curve $M=M_{PPDL}$ ($M=M_{NPDL}$), one can get a PPDL (NPDL) solution.
  \item In absence of $M_{NPDL}$ ($ M_{max}$), $ M_{max}$ ($ M_{NPDL}$) is the upper bound of $ M $ for the existence of NPSWs, i.e., there does not exist any NPSW if $ M > M_{max}$ ($ M> M_{NPDL}$). Although, it is important to note that there exist NPSWs along the curve $M=M_{max}$ but there does not exist any NPSW along the curve $M=M_{NPDL}$.
  \item If we pick a $\beta_{i}$ and go vertically upwards, then all intermediate values of $ M $ bounded by the curves $ M=M_{c} $ and $ M=M_{max} $ or $M_{NPDL}$ or $\max\{M_{max},M_{NPDL}\}$ would give NPSWs.
  \item Similarly, all intermediate values of $ M $ bounded by the curves $ M=M_{c} $ and $ M=M_{PPDL} $ would give PPSWs.
  \item We have used the following notations to interpret the different solution spaces  : C -- Region of coexistence of both NPSWs and PPSWs, N -- Region of existence of NPSWs and P -- Region of existence of PPSWs.
\end{itemize}

\begin{figure}
\begin{center}
  \includegraphics{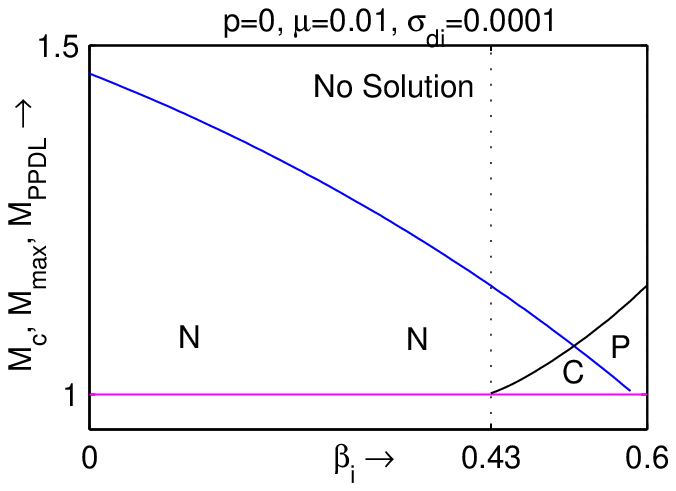}
  \caption{\label{sol_spc_wrt_beta_i_p=0} Existence domain with respect to $\beta_{i}$ for $p=0$, $\mu=0.01$, $\sigma_{di}=0.0001$, $\sigma_{ie}=0.9$ and $\sigma_{ip}=1$. The black curve, the magenta curve and the blue curve correspond to the curves $M=M_{PPDL}$, $M=M_{c}$ and $M=M_{max}$ respectively.}
\end{center}
\end{figure}

\begin{figure}
\begin{center}
  \includegraphics{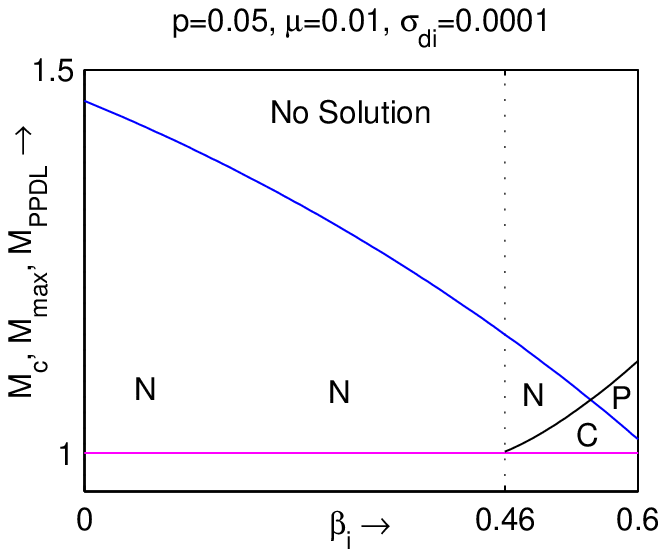}
  \caption{\label{sol_spc_wrt_beta_i_p=0_pt_05} Existence domain with respect to $\beta_{i}$ for $p=0.05$, $\mu=0.01$, $\sigma_{di}=0.0001$, $\sigma_{ie}=0.9$ and $\sigma_{ip}=1$. The black curve, the magenta curve and the blue curve correspond to the curves $M=M_{PPDL}$, $M=M_{c}$ and $M=M_{max}$ respectively.}
\end{center}
\end{figure}

\begin{figure}
\begin{center}
  \includegraphics{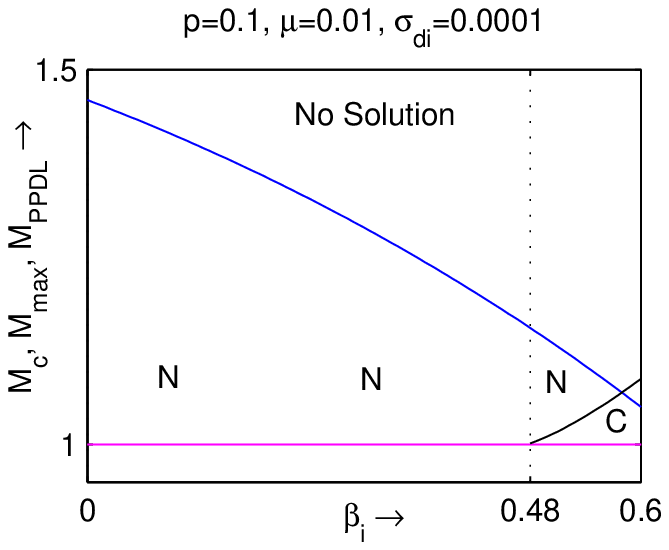}
  \caption{\label{sol_spc_wrt_beta_i_p=0_pt_1} Existence domain with respect to $\beta_{i}$ for $p=0.1$, $\mu=0.01$, $\sigma_{di}=0.0001$, $\sigma_{ie}=0.9$ and $\sigma_{ip}=1$. The black curve, the magenta curve and the blue curve correspond to the curves $M=M_{PPDL}$, $M=M_{c}$ and $M=M_{max}$ respectively.}
\end{center}
\end{figure}

\begin{figure}
\begin{center}
\includegraphics{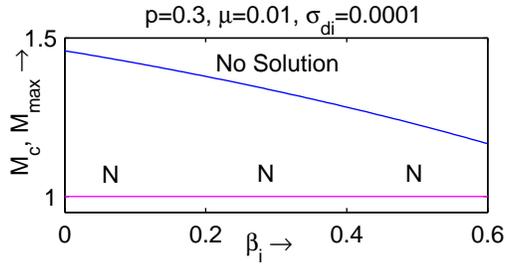}
  \caption{\label{sol_spc_wrt_beta_i_p=0_pt_3} Existence domain with respect to $\beta_{i}$ for $p=0.3$, $\mu=0.01$, $\sigma_{di}=0.0001$, $\sigma_{ie}=0.9$ and $\sigma_{ip}=1$. The magenta curve and the blue curve correspond to the curves $M=M_{c}$ and $M=M_{max}$ respectively.}
\end{center}
\end{figure}

\begin{figure}
\begin{center}
\includegraphics{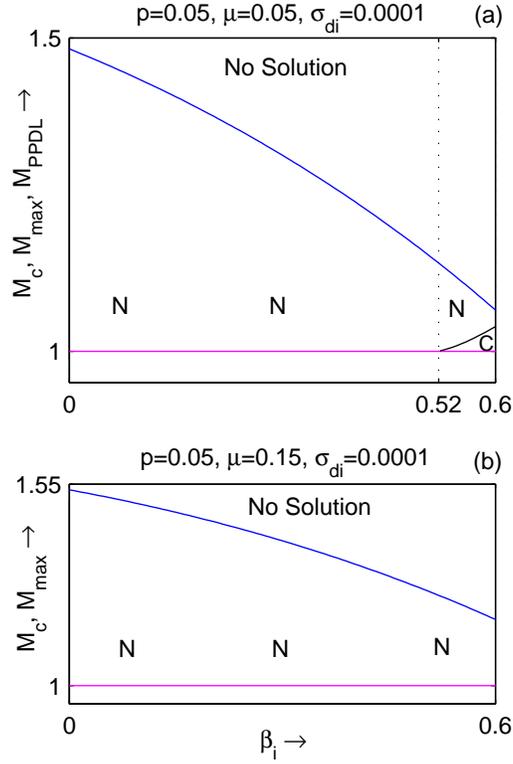}
  \caption{\label{final_sol_spc_wrt_beta_i_for_diff_mu} Existence domain with respect to $\beta_{i}$ with $p=0.05$, $\sigma_{di}=0.0001$, $\sigma_{ie}=0.9$ and $\sigma_{ip}=1$ for (a) $\mu=0.05$ and (b) $\mu=0.15$. In both figures, the black curve, the magenta curve and the blue curve correspond to the curves $M=M_{PPDL}$, $M=M_{c}$ and $M=M_{max}$ respectively.}
\end{center}
\end{figure}

Figures \ref{sol_spc_wrt_beta_i_p=0} - \ref{sol_spc_wrt_beta_i_p=0_pt_3} are the existence domains with respect to $\beta_{i}$ for different values of $p$ starting from $p=0$ whenever $\mu=0.01$, $\sigma_{di}=0.0001$ and $\sigma_{ie}=\sigma_{pe}=0.9$. These figures show the nature of existence of different DA solitary structures with the variation of positron concentration ($p$) in the system. Figure \ref{final_sol_spc_wrt_beta_i_for_diff_mu}(a) and \ref{final_sol_spc_wrt_beta_i_for_diff_mu}(b) are the existence domains with respect to $\beta_{i}$ for $\mu=0.05$ and $\mu=0.15$, respectively, whenever $p=0.05$, $\sigma_{di}=0.0001$, and $\sigma_{ie}=\sigma_{pe}=0.9$.  With respect to the above mentioned general description of the existence domains these figures are self explanatory. For example, let us consider figure \ref{sol_spc_wrt_beta_i_p=0_pt_05} which shows the existence domain with respect to $\beta_{i}$ for $p=0.05$, $\mu=0.01$, $\sigma_{di}=0.0001$, and $\sigma_{ie}=\sigma_{pe}=0.9$. From this figure we observe the following:
\begin{itemize}
	\item The system supports NPSWs whenever $\beta_{i}$ lies within the interval $0 \leq \beta_{i} < \beta_{i}^{(b)}$ and the existence region of NPSWs is bounded by the curves $M=M_{c}$ and $M=M_{max}$.
	\item For $\beta_{i} > \beta_{i}^{(a)}$, the system starts to support PPDLs along the curve $M=M_{PPDL}$.
	\item In the interval $\beta_{i}^{(a)} < \beta_{i} < 0.6$, the system supports PPSWs and the existence region of PPSWs is bounded by the curves $M=M_{c}$ and $M=M_{max}$.
	\item As the curve $M=M_{PPDL}$ restricts the existence of PPSWs, there does not exist PPSWs after the formation of PPDL. Consequently, the system does not support DA supersoliton of positive polarity.
	\item In the interval $\beta_{i}^{(a)} < \beta_{i} < \beta_{i}^{(b)}$, the system supports coexistence of solitary waves of both polarities. The coexistence region of PPSWs and NPSWs is bounded by the curves $M=M_{c}$, $M=M_{max}$ and $M=M_{PPDL}$. For $p=0.05$, $\mu=0.01$, $\sigma_{di}=0.0001$, and $\sigma_{ie}=\sigma_{pe}=0.9$, the values of $\beta_{i}^{(a)}$ and $\beta_{i}^{(b)}$ are 0.46 and 0.6 respectively.
\end{itemize}

Now figure \ref{sol_spc_wrt_beta_i_p=0} shows that whenever there is no positron in the system it supports NPSWs, PPSWs, co-existence of solitary waves of both polarities and PPDLs. These observations agree with the observations of Das \textit{et al.} \cite{das12mc} for supersonic DA solitary structures in a three component electron-ion-dust plasma consisting of Cairns \cite{cairns95} distributed nonthermal ions, isothermal electrons and adiabatic warm dust grains. Now, if we inject an amount of isothermal positrons in the system, then up to a cut-off value $p^{(c)}$ of positron concentration $p$, we do not observe any qualitative change in the existence domains (see figure \ref{sol_spc_wrt_beta_i_p=0_pt_05}). But in the interval $0<p<p^{(c)}$, the existence regions of PPDL and PPSW decrease for increasing values of $p$ (see figures \ref{sol_spc_wrt_beta_i_p=0_pt_05} and \ref{sol_spc_wrt_beta_i_p=0_pt_1}). Finally, for $p>p^{(c)}$ the system does not support positive potential solitary structures any more (see figure \ref{sol_spc_wrt_beta_i_p=0_pt_3}). It only supports NPSWs in a region bounded by the curves $M=M_{c}$ and $M=M_{max}$ for the entire range of $\beta_{i}$.

Again, figures \ref{sol_spc_wrt_beta_i_p=0_pt_05}, \ref{final_sol_spc_wrt_beta_i_for_diff_mu}(a) and \ref{final_sol_spc_wrt_beta_i_for_diff_mu}(b) show the existence domains with respect to $\beta_{i}$ for $\mu=0.01$, $\mu=0.05$ and $\mu=0.15$, respectively, whenever $p=0.05$, $\sigma_{di}=0.0001$, and $\sigma_{ie}=\sigma_{pe}=0.9$. From these figures we see that for a fixed positron concentration, if we increase the values of $\mu$ from $\mu=0.05$ then the existence regions of both PPDL and PPSW decrease. Finally, there exists a cut-off value $\mu^{(c)}$ of $\mu$ such that for $\mu>\mu^{(c)}$ the system does not support any positive potential solitary structure for any physically admissible value of $\beta_{i}$. For $\mu>\mu^{(c)}$, the system only supports NPSWs for the entire range of $\beta_{i}$ and the existence region of NPSWs is bounded by the curves $M=M_{c}$ and $M=M_{max}$ (see figure \ref{final_sol_spc_wrt_beta_i_for_diff_mu}(b)).

\begin{figure}
\begin{center}
\includegraphics{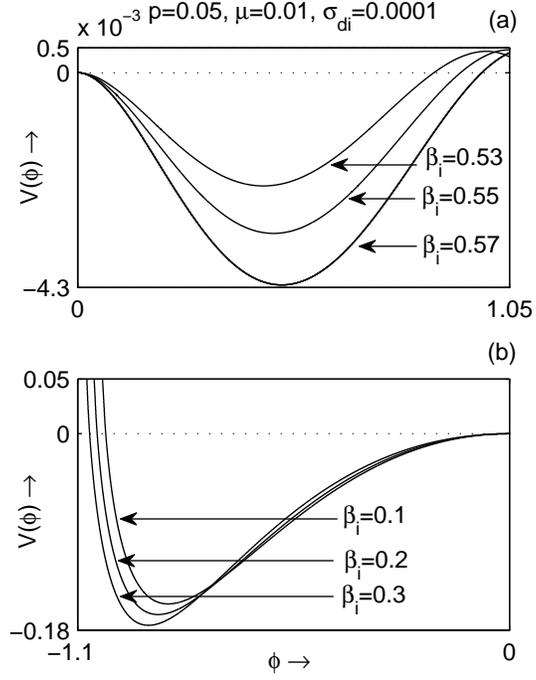}
  \caption{\label{final_phi_v_phi_for_diff_beta_i} $V(\phi)$ is plotted against $\phi$ for different values of $\beta_{i}$ and fixed values of other parameters: (a) $p=0.05$, $\mu=0.01$, $\sigma_{di}=0.0001$, $\sigma_{ie}=0.9$ and $\sigma_{ip}=1$ and (b) $p=0.05$, $\mu=0.01$, $\sigma_{di}=0.0001$, $\sigma_{ie}=0.9$ and $\sigma_{ip}=1$.}
\end{center}
\end{figure}

\begin{figure}
\begin{center}
  \includegraphics{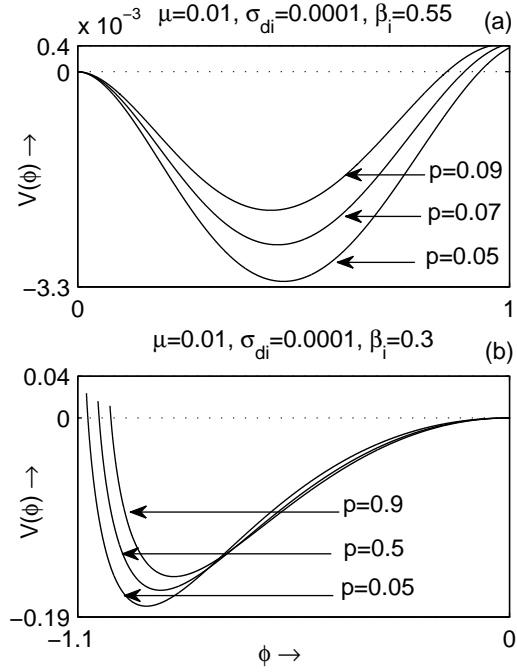}
  \caption{\label{final_phi_v_phi_for_diff_p} $V(\phi)$ is plotted against $\phi$ for different values of $p$ and fixed values of other parameters: (a) $\beta_{i}=0.55$, $\mu=0.01$, $\sigma_{di}=0.0001$, $\sigma_{ie}=0.9$ and $\sigma_{ip}=1$ and (b) $\beta_{i}=0.3$, $\mu=0.01$, $\sigma_{di}=0.0001$, $\sigma_{ie}=0.9$ and $\sigma_{ip}=1$. }
\end{center}
\end{figure}

\begin{figure}
\begin{center}
  \includegraphics{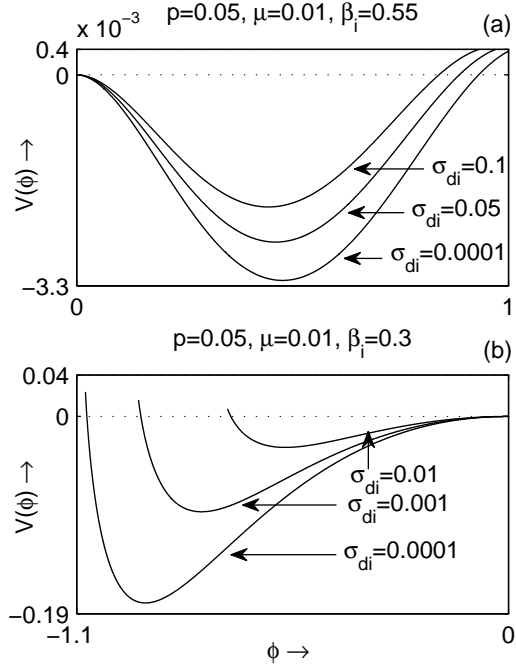}
  \caption{\label{final_phi_v_phi_for_diff_sigma_di} $V(\phi)$ is plotted against $\phi$ for different values of $p$ and fixed values of other parameters: (a) $p=0.05$, $\beta_{i}=0.55$, $\mu=0.01$, $\sigma_{ie}=0.9$ and $\sigma_{ip}=1$ and (b) $p=0.05$, $\beta_{i}=0.3$, $\mu=0.01$, $\sigma_{ie}=0.9$ and $\sigma_{ip}=1$.}
\end{center}
\end{figure}

We have the following observations regarding the amplitude of PPSWs and NPSWs:
\begin{itemize}
\item The amplitude of PPSW increases with increasing values of $\beta_{i}$ (see figure \ref{final_phi_v_phi_for_diff_beta_i}(a)) for fixed values of 
the other parameters of the present plasma model.
\item The amplitude of NPSW increases with increasing values of $\beta_{i}$ (see figure \ref{final_phi_v_phi_for_diff_beta_i}(b)) for fixed values of the other parameters of the present plasma model.
\item The amplitude of PPSW decreases with increasing values of $p$ (see figure \ref{final_phi_v_phi_for_diff_p}(a)) for fixed values of the other parameters of the present plasma model.
\item The amplitude of NPSW decreases with increasing values of $p$ (see figure \ref{final_phi_v_phi_for_diff_p}(b)) for fixed values of the other parameters of the present plasma model.
\item The amplitude of PPSW decreases with increasing values of $\sigma_{di}$ (see figure \ref{final_phi_v_phi_for_diff_sigma_di}(a)) for fixed values of the other parameters of the present plasma model.
\item The amplitude of NPSW decreases with increasing values of $\sigma_{di}$ (see figure \ref{final_phi_v_phi_for_diff_sigma_di}(b)) for fixed values of the other parameters of the present plasma model.
\end{itemize}

\section{Phase Portraits of Solitary Structures}\label{sec:Phase_Portraits_c5}

From the discussions of previous section we have seen that the present system supports following DA solitary structures: NPSWs, PPSWs, coexistence of solitary waves of both polarities and PPDLs. We now present the phase portraits of the dynamical system corresponding to the above mentioned DA solitary structures.

Differentiating the energy integral (\ref{energy_integral}) with respect to $\phi$, we get the following differential equation:
\begin{eqnarray}\label{energy_integral_differentiation}
\frac{d^{2}\phi}{d\xi^{2}}+V'(\phi)=0.
\end{eqnarray}
This equation is equivalent to the following system of coupled differential equations
\begin{eqnarray}\label{phase_portraits_c5}
\frac{d\phi_{1}}{d\xi}=\phi_{2}~,~\frac{d\phi_{2}}{d\xi}=-V'(\phi_{1})~,
\end{eqnarray}
where $\phi_{1}=\phi$. Now, we shall confirm the existence of the DA solitary structures with the help of phase portraits of the system of coupled equations (\ref{phase_portraits_c5}) in the $\phi_{1}-\phi_{2}$ plane.

\begin{figure}
\begin{center}
\includegraphics{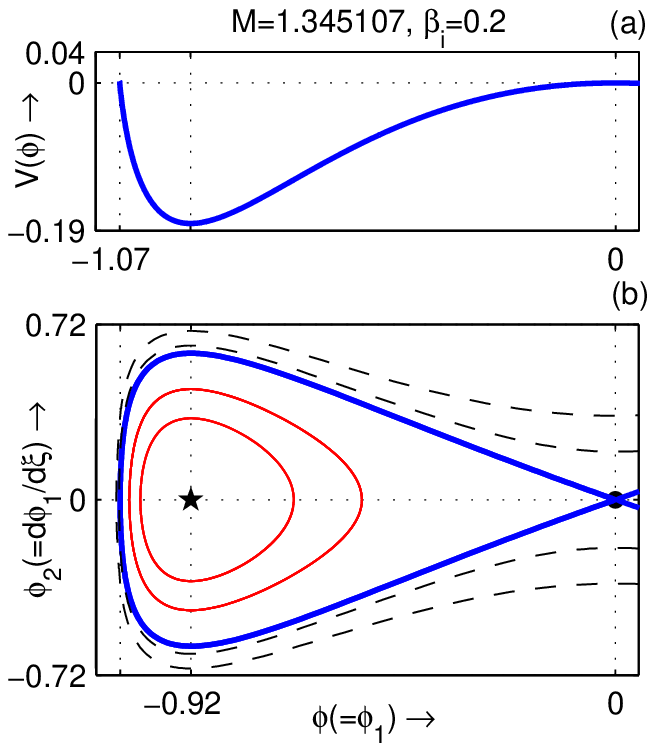}
  \caption{\label{final_pp_NPSW} $V(\phi)$ (on top) and the phase portrait of the system (\ref{phase_portraits_c5}) (on bottom) have been drawn on the same $\phi(=\phi_{1})$-axis at $M=M_{c}$ when $p=0.05$, $\mu=0.01$, $\beta_{i}=0.3$, $\sigma_{ie}=0.9$ and $\sigma_{ip}=1$.}
\end{center}
\end{figure}

\begin{figure}
\begin{center}
\includegraphics{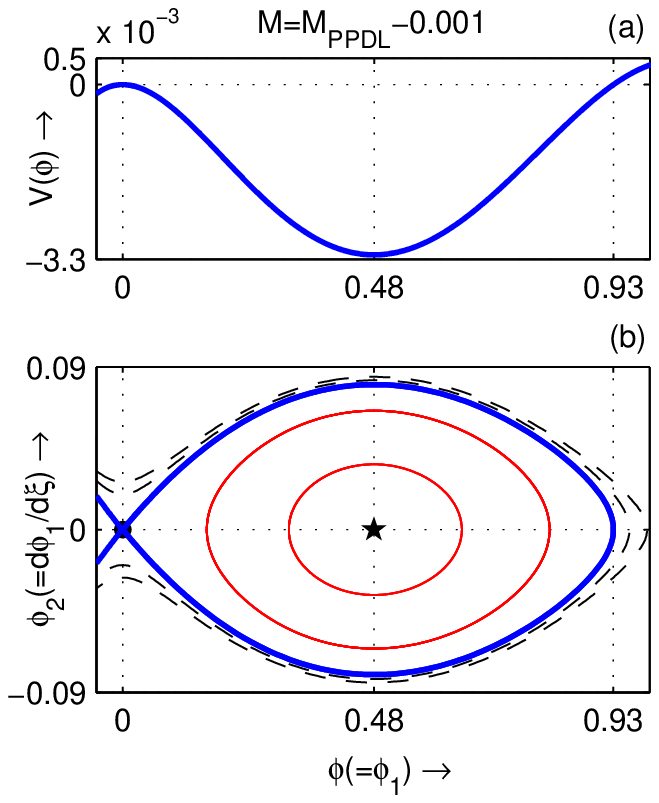}
  \caption{\label{final_pp_PPSW} $V(\phi)$ (on top) and the phase portrait of the system (\ref{phase_portraits_c5}) (on bottom) have been drawn on the same $\phi(=\phi_{1})$-axis at $M=M_{c}$ when $p=0.05$, $\mu=0.01$, $\beta_{i}=0.55$, $\sigma_{ie}=0.9$ and $\sigma_{ip}=1$.}
\end{center}
\end{figure}

\begin{figure}
\begin{center}
\includegraphics{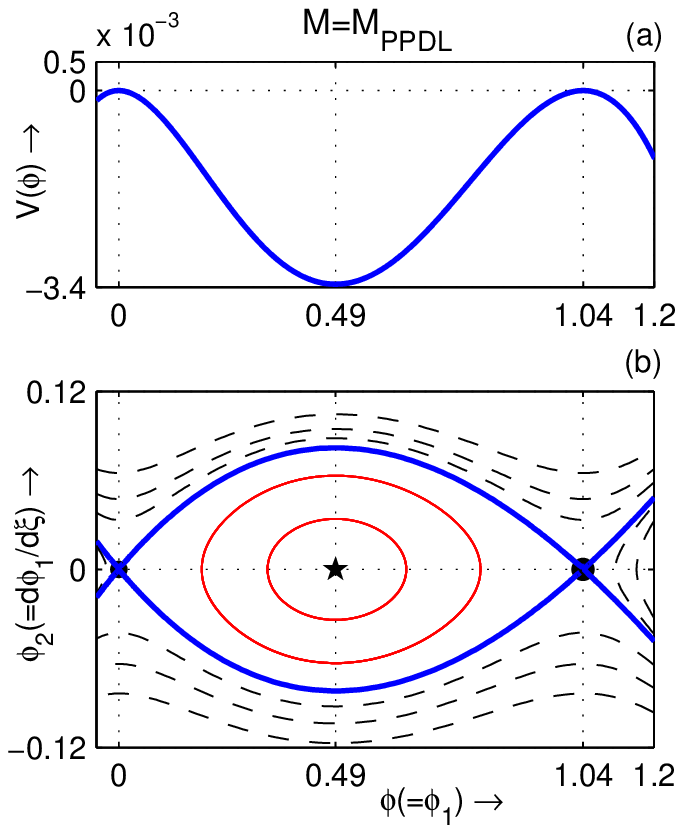}
  \caption{\label{final_pp_PPDL} $V(\phi)$ (on top) and the phase portrait of the system (\ref{phase_portraits_c5}) (on bottom) have been drawn on the same $\phi(=\phi_{1})$-axis at $M=M_{c}$ when $p=0.05$, $\mu=0.01$, $\beta_{i}=0.55$, $\sigma_{ie}=0.9$ and $\sigma_{ip}=1$.}
\end{center}
\end{figure}

\begin{figure}
\begin{center}
\includegraphics{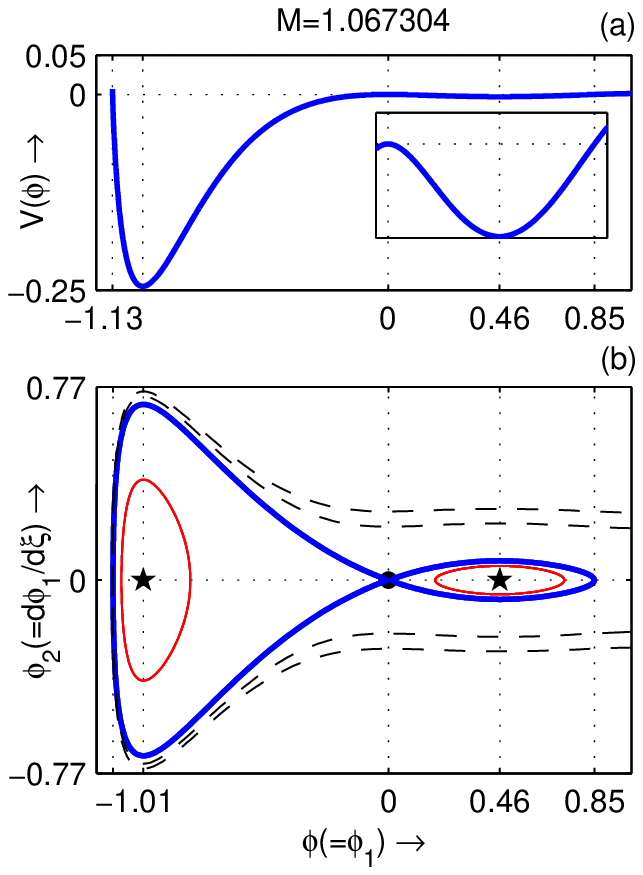}
  \caption{\label{final_pp_coexistence} $V(\phi)$ (on top) and the phase portrait of the system (\ref{phase_portraits_c5}) (on bottom) have been drawn on the same $\phi(=\phi_{1})$-axis at $M=M_{c}$ when $p=0.05$, $\mu=0.01$, $\beta_{i}=0.5$, $\sigma_{ie}=0.9$ and $\sigma_{ip}=1$.}
\end{center}
\end{figure}

For this purpose, we consider figures \ref{final_pp_NPSW} - \ref{final_pp_coexistence}. Here, we have used the existence domain as shown in figure \ref{sol_spc_wrt_beta_i_p=0_pt_05} to determine the value of $M$ for the existence of desired solitary structure. In each figure of figures \ref{final_pp_NPSW}(a) - \ref{final_pp_coexistence}(a), $V(\phi)$ is plotted against $\phi$. The lower panel (or marked as (b)) of each figure shows the phase portrait of the system (\ref{phase_portraits_c5}). In these figures, we have used the values of the parameters as indicated in the figures with $p=0.05$, $\mu=0.01$, $\sigma_{di}=0.0001$, and $\sigma_{ie}=\sigma_{pe}=0.9$. The curve $V(\phi)$ and the  phase portrait have been drawn on the same horizontal axis $\phi(=\phi_{1})$. The small solid circle corresponds to a \textbf{saddle point (unstable equilibrium point)} and the small solid star indicates an equilibrium point other than saddle point (\textbf{stable equilibrium point}). 

Now, there is a one-one correspondence between the separatrix of the phase portrait as shown with a heavy blue line in the lower panel with the curve $V(\phi)$ against $\phi$ of the upper panel. In fact, this one-one correspondence between the separatrix of the phase portrait and the curve $V(\phi)$ against $\phi$ has been elaborately discussed in the previous chapters for DIA solitary structures. In this section of the present paper, we like to show the phase portraits of the DA solitary structures supported by the system. For the easy readability of this section let us recall some fundamental discussions about the one-one correspondence between the separatrix of the phase portrait with the curve $V(\phi)$ against $\phi$ as explained in the paper of Paul \textit{et al.} \cite{paul17pop} 

Since each maximum point of the curve $V(\phi)$ correspond to a saddle point of the system, the origin $(0,0)$ must be a saddle point of the system (\ref{phase_portraits_c5}) and the separatrix corresponding to a solitary structure (shown by a heavy blue line) appears to pass through the saddle at the origin and encloses the stable equilibrium points which correspond to the minimum points of the curve $V(\phi)$. This separatrix encloses infinitely many closed curves (shown by red lines) about a stable equilibrium point. This closed curves indicate the existence of periodic wave solutions. Here, it is important to note that, the separatrix corresponding to a solitary structure is not a closed curve and therefore, the separatrix does not indicate the periodic wave solution. Because for this case, the trajectory takes forever trying to reach a saddle point. In fact, this is the reason that a pseudo-particle associated with the energy integral (\ref{energy_integral}) takes an infinite long time to move away from its unstable position of equilibrium and then it continues its motion until $\phi$ takes the value $\phi_{m} (>0)$, where $V(\phi_{m})=0$ and $V'(\phi_{m})>0$ and again it takes an infinite long time to come back its unstable position of equilibrium \cite{verheest00}.

Figure \ref{final_pp_NPSW}(a) confirms the existence of a NPSW and figure \ref{final_pp_NPSW}(b) shows that the corresponding phase portrait contains only one saddle at the origin and a non-zero stable equilibrium point at $(-0.92,0)$. Consequently, there exists only one separatrix on the negative side of $\phi_{1}$ axis and the separatrix appears to start and end at the origin enclosing the stable equilibrium point $(-0.92,0)$. Figure \ref{final_pp_PPSW}(b) shows the phase portrait of a PPSW corresponding to the PPSW shown in figure \ref{final_pp_PPSW}(a). Here we see that there exist one saddle at the origin and a non-zero stable equilibrium point at $(0.48,0)$. So only one separatrix is possible on the side of negative $\phi_{1}$ axis and the separatrix appears to start and end at the origin enclosing the stable equilibrium point $(0.48,0)$. Figure \ref{final_pp_PPDL}(a) shows the existence of a PPDL and figure \ref{final_pp_PPDL}(b) describes the corresponding phase portrait. This figure shows that the separatrix corresponding to the double layer solution appears to start and end at the saddle $(0,0)$ and again it appears to pass through the saddle point at $(1.04,0)$ enclosing the stable equilibrium point $(0.49,0)$. Figure \ref{final_pp_coexistence}(a) shows the coexistence of solitary waves of both polarities and figure \ref{final_pp_NPSW}(b) shows that the corresponding phase portrait. Here the separatrix appears to pass through the saddle at the origin enclosing both the stable equilibrium points $(-1.01,0)$ and $(0.46,0)$.


\section{Conclusions}\label{sec:conclusions_c5}

In the present work, we have investigated the nature of existence of different DA solitary structures in a collisionless unmagnetized dusty plasma consisting of negatively charged warm dust grains, Cairns \cite{cairns95} distributed nonthermal ions, isothermal electrons and isothermal positrons with the help of existence domains and phase portraits. We have observed the following:
\begin{itemize}
	\item The system supports NPSWs, PPSWs, PPDLs and coexistence of solitary waves of both polarities.
	\item The positive potential double layer solution of the energy integral restricts the occurrence of PPSWs. Consequently, there does not exist PPSW after the formation of double layer. So the system does not support positive potential supersoliton.
	\item The system does not support NPDLs. So there is no question of existence of negative potential supersolitons.
	\item With the help of the existence domains we see that the existence of PPDLs and PPSWs depend on the concentration of positrons in the system. In fact, for increasing positron concentration the existence region of PPDLs and PPSWs decrease and ultimately both of them vanish from the system.
	\item The existence of PPDLs and PPSWs also depend on the concentration of electrons in the system. The nonthermal parameter of ions $\beta_{i}$ has an important role on the existence of DA double layers of positive polarity. For isothermal ions the system does not supports PPDL, rather PPDLs are observed for strong nonthermality of ions.
	\item We have also studied the variation of amplitudes of NPSWs and PPSWs with respect to the parameters $\beta_{i}$, $p$ and $\sigma_{di}$.
	\item Finally, we have presented the phase portraits of the dynamical system to confirm the existence of DA solitary structures. 
\end{itemize}

\acknowledgments One of the authors (Ashesh Paul) is thankful to the Department of Science and Technology, Govt. of India, INSPIRE Fellowship Scheme for financial support.


\providecommand{\noopsort}[1]{}\providecommand{\singleletter}[1]{#1}%

\newpage


\end{document}